\documentclass[aps,reprint,pra, twocolumn,superscriptaddress,floats,floatfix]{revtex4-1}

\usepackage{amsmath}
\usepackage{amsfonts}
\usepackage{amssymb}
\usepackage{graphicx}
\usepackage{color}
\usepackage{tikz}
\usepackage[colorlinks=true,citecolor=blue,linkcolor=blue,urlcolor=blue]{hyperref}
\usepackage{times}
\usepackage{amstext}
\usepackage{latexsym}
\usepackage[running,mathlines]{lineno}
\usepackage{bbm}
\usepackage{verbatim}

\providecommand{\ket}[1]{\lvert #1 \rangle}

\usepackage{orcidlink}

\newcommand{\UFSCar}{Departamento de Física, Universidade Federal de São Carlos, PO box 676, 13565-905 São Carlos, SP, Brazil}

\newcommand{\IFSC}{São Carlos Institute of Physics, University of São Paulo, PO Box 369, 13560-970, São Carlos, SP,
Brazil}

\begin{document}
\raggedbottom

\title{Pulsed Laser as a Continuous Particle Stream}

\author{Ciro Micheletti~Diniz~\orcidlink{0000-0002-7602-0468}}
\email{mdciro@df.ufscar.br }
\affiliation{\UFSCar}

\author{Franciele Renata Henrique~\orcidlink{0009-0000-1525-6314}}
\affiliation{\UFSCar}

\author{Bruno Santos de Souza~\orcidlink{0009-0003-9346-2844}}
\affiliation{\UFSCar}

\author{Lino Misoguti~\orcidlink{0000-0001-6624-8453}}
\affiliation{\IFSC}

\author{Paulo Henrique Dias Ferreira~\orcidlink{0000-0003-1694-2126}}
\affiliation{\UFSCar}

\author{Celso Jorge Villas~Boas~\orcidlink{0000-0001-5622-786X}}
\email{celsovb@df.ufscar.br}
\affiliation{\UFSCar}

\date{\today}

\begin{abstract}

With the recently introduced particle interpretation of the double-slit experiment for light fields [Phys. Rev. Lett. \textbf{134}, 13360 (2025)], all related interference phenomena can be reinterpreted in terms of light particle states that either couple (bright) or do not couple (dark) with detectors. Here, we apply this approach to multimode pulsed lasers, unifying the description of their generation inside optical cavities and propagation outside them, now relying solely on quantum mechanics, i.e., without invoking wave superposition to explain pulse shaping. Specifically, we demonstrate that multimode interference presents a dominance of particle dark states over the bright ones and mode-locked pulsed lasers consist of a continuous photon beam, with photons forming bright states during pulses and dark states in between. Additionally, we analyze mode-locked pulsed lasers, showing that the theoretical bright-to-dark state ratio matches the experimental pulse-to-interval duration ratio.

\end{abstract}

\maketitle

Over the past decades, lasers have become essential tools in fields such as medicine, communications, manufacturing, and fundamental science~\cite{Editorial_lasers, Review_laser_solid_state, Review_laser_cancer, Review_laser_quantum_comunication, Review_laser_mining}. The extension of masers operating in infrared and optical regions~\cite{laser_formulation1} and the following experimental demonstration by Maiman in 1960~\cite{maiman1960stimulated} marked a technological breakthrough, driving advances in areas like nonlinear optics~\cite{franken1961generation}, laser processing~\cite{Salter2019}, and classical~\cite{Miller2012} and quantum~\cite{Gisin2007} communications. Lasers have also deepened our understanding of nature, enabling femtochemistry~\cite{zewail1988laser}, attosecond physics~\cite{hentschel2001attosecond}, and groundbreaking discoveries like those made by LIGO~\cite{Abbott2009}.

Lasers operate based on the excitation of gain media inside optical cavities and the stimulated emission of photons~\cite{einstein1917}. While commonly available laser devices operate in continuous wave (CW) mode, pulsed lasers are of great interest and widely used in industry and research. In fact, the very first laser operated in a pulsed regime~\cite{maiman1960stimulated}. For many gain media, including ruby used by Maiman, pulsed operation is the only possibility, as their three-level system can achieve population inversion exclusively through strong pumping, typically provided by flash or arc lamps. Pulsed laser operation can also be achieved through other mechanisms, such as Q-switching~\cite{mcclung1962giant} and mode-locking~\cite{hargrove1964locking,mocker1965mode}, enabling the generation of ultrashort laser pulses with temporal width on the order of femtoseconds~\cite{keller2021ultrafast}, or even attoseconds~\cite{strickland1985compression,antoine1996attosecond}. These pulses are characterized by their high peak intensity and broad emission spectra~\cite{strickland1985compression,antoine1996attosecond}.

Inside optical cavities, where the gain material is continuously emitting photons with different energies, the stimulated emission~\cite{einstein1917} and the atomic population inversion are precisely described by the quantum treatment \cite{Walls_and_Milburn_1ed, scully67, book_laser_fundamentals_applicatons}, although some semiclassical description may be possible~\cite{Lamb64, Stenholm69, Allen_69, shirley69, book_laser_fundamentals_applicatons}. However, the behavior and propagation of pulsed lasers (outside optical cavities) must be derived from Maxwell's wave equation, for instance, by considering the interference of allowed cavity modes when their amplitudes and phases are constant, in the case of mode-locking. Although wave interpretation perfectly matches the pulse stream in mode-locked lasers, difficulty arises when trying to describe what happens with the continuous emission of photons by the gain material, i.e., ``How can continuously emitted photons remain undetectable in certain (dark) regions or instances of time?'' In this letter, we intend to solve this puzzle by employing the recent reinterpretation of wave interference phenomena in terms of collective superposition particle (photon) states that either couple (bright states -- BS) or do not couple (dark states -- DS) with matter~\cite{Celso_DBS}. 

Considering a mode-locked pulsed laser, we demonstrate that a continuous stream of photons exists either in bright states (pulse interval) or dark states (interval between pulses), thus introducing a unified particle explanation for laser operation, i.e., the laser emission inside cavities and its shaping and propagation outside them. To this end, we extend the previous analyses of interference in double-slit experiments in terms of bright and dark states~\cite{Celso_DBS} to multimode interference. Although such generalization is mathematically straightforward, it reveals the emergence of a new physical phenomenon: unlike the two-mode case, where bright and dark states always appear in pairs for any $N$-photon state, the multimode scenario is characterized by the presence of a significantly higher number of dark (undetectable) states per bright (detectable) state, which increases with the number of superposed modes and the total number of photons. This is precisely the case we find either in multislit experiments, Fig.~\ref{Fig:Model}(a), or, analogously, in mode-locked pulsed lasers, Fig.~\ref{Fig:Model}(b): in both cases the phase relation between modes is constant and implies a number of dark states much higher than the number of bright states, thus describing long intervals with undetectable photons (dark intervals) and short intervals of detectable photons (pulse intervals)~\footnote{In multislit experiments, interference arises from the sum of different spatial modes (with the same frequency), each with time-independent phases determined by their respective paths to the screen.  On the other hand, in pulsed lasers, the interference arises from the sum of different modes with distinct frequencies, which results in time-dependent phases.}. To describe a pulsed laser solely in terms of a particle beam, we must first review its standard description and properties within a fully wave-based framework. Later, we will explore its similarity to multislit experiments and then describe both using a fully quantum model. 

Pulsed lasers originating from the mode-locking technique are based on the interference of longitudinal modes of the electric field inside a cavity in such a way that the amplitude of the resulting field can be written as the summation of all modes, as follows~\cite{book_laser_fundamentals_applicatons}
\begin{equation}\label{eq1}
E(t) = \sum^{M}_{m=1}{E_m(t)e^{i\left[\omega_m t+\varphi_m(t)\right]}},
\end{equation}
\noindent where $M$ is the number of modes. $E_m(t)$, $\varphi_m(t)$, and $\omega_m = \omega_0 + [m-(M+1)/2]\Delta\omega$~\footnote{Assuming M odd for simplicity.} are, respectively, the amplitude, the time-dependent phase, and the frequency of the $m$-th mode, with $\omega_0$ being the central frequency, and $\Delta\omega =\frac{\pi c}{n_r L}$ being the free spectral range for a Fabry-Perot cavity, where $n_r$ is the refractive index of the cavity medium, $L$ is the length of the cavity, and $c$ is the speed of light in vacuum. For $M\gg1$, the locked time-dependent phases follow the linear relation $\varphi_m(t)-\varphi_{m-1}(t) = \varphi_0$ (constant) and, as $\omega_0\gg \Delta \omega$, we can assume all $E_m(t)$ to be constant and equal to $E_0$~\cite{svelto2010principles}. Thus, we can rewrite Eq.~\eqref{eq1} as $E(t)=A(t)e^{i[\omega_0 - (M+1)\Delta\omega/2]t}$ with
\begin{equation}\label{eq2}
A(t) = \sum_{m=1}^{M}{E_0e^{im(\Delta\omega t+\varphi_0)}}.
\end{equation}
Then, by changing to another time reference in which $\Delta\omega t'=\Delta\omega t+\varphi_0$, the summation can be expressed as a geometric progression, resulting in
\begin{equation}\label{eq3}
A(t') \propto  E_0\text{sin}\left[M\Delta\omega t'/2\right]/\text{sin}\left(\Delta\omega t'/2\right). 
\end{equation}
\noindent Since the laser beam intensity is proportional to $A^2(t')$, one can conclude that the interference of different modes of a laser cavity, when their phases are locked, results in a train of pulses, separated by the cavity's flying time $\tau=2\pi/\Delta\omega$~\cite{svelto2010principles}. Additionally, the higher the value of M, the higher the peak intensity and the shorter the pulse duration.

Now, to introduce the particle explanation of the train of pulses, firstly, we will recall the multislit experiments to make a clear comparison with the results in~\cite{Celso_DBS}. As detailed in the Supplementary Material (SM), which includes the Refs.~\cite{book_laser_fundamentals_applicatons, svelto2010principles, Walls_and_Milburn_1ed, Double_slit_time_interference, double_slit_interference_geometry, Celso_DBS, book_complex_analysis, Scully_Zubairy_1997, Zou2020, Yang_25, mode_lock_optic_fiber_PRL, Wang_12_TiSa_laser, mode_lock_random_laser_nature_communi}, a stream of photons passing through a diffraction grating with $M$ slits can be described by a state with phases $m[kd\sin(\theta)]$~\cite{Scully_Zubairy_1997}, where $d$ is the slit separation, $k =|\vec{k}_m|$ with $\vec{k}_m$ being the wave vector of the $m$-th mode (slit), and $\theta$ is the angle between the horizontal axis and the position vector $\vec{r}$ on a given reference point on the screen~\cite{double_slit_interference_geometry} -- see Fig.~\ref{Fig:Model}(a). Thus, the phases resulting from the sum of multiple modes arising after light passes through a diffraction grating~\cite{Yashin98, Bonod16, Palmer20, sinha2010ruling, magana2016exotic} are exactly analogous to the ones present in Eq.~\eqref{eq2}. By leveraging this analogy, the behavior of a pulsed laser can be modeled through the same analytical framework used for multislit diffraction~\cite{Double_slit_time_interference}. Moreover, a comprehensive analysis can be done by disregarding the physical constraints of the problems and looking only at the mathematical description of a general case of multimodal interference.

\begin{figure}[t!]
\includegraphics[width=1.0\linewidth]{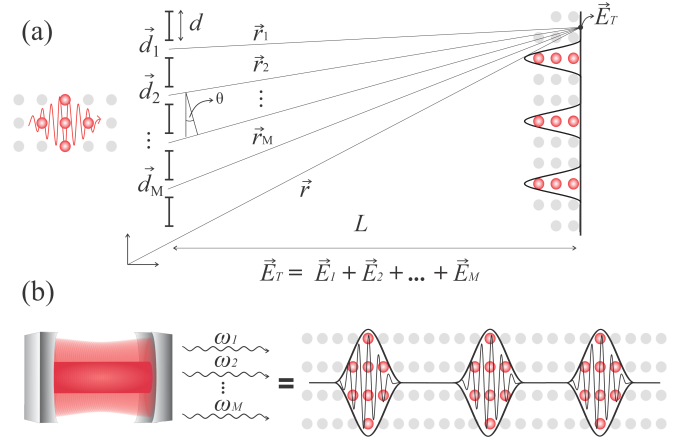}
	\caption{(a) Illustration of the diffraction pattern resulting from coherent light impinging on a grating with $M$ slits. The distance between the diffraction grating and the screen is $L$, the separation of the slits is $d$, and the position of the $m$-th slit is $\vec{d}_m$, with $1 \leq m \leq M$. The distance between a reference point on the grating and a reference point on the screen is given by $\vec{r}$, and the vector connecting the $m$-th slit to the reference point on the screen is given by $\vec{r}_m$. The electric field operator (positive frequency) in the reference point on the screen, located at $\vec{r}$, is given by (apart from a constant factor) $\vec{E}_T ^{(+)}= \sum_{m=1}^M \vec{E}_m^{(+)}$, where $\vec{E}_m ^{(+)}= E_m^{(+)}\vec{e_m}$ ($\vec{e_m}$ is the direction of the electric field), with $E_m^{(+)}= a_me^{i\vec{k}_m\cdot\vec{r}}$. (b) Pictographic representation of the pulsed laser photons. The gray (red) circles represent the photons in the regions where the collective state for the beam is a dark (bright) state or a combination of dark (intermediate) states, following the notation from Ref. \cite{Celso_DBS}. As depicted, in contrast to what is observed in the bright regions, the detectable beam intensity in the dark regions is zero despite the presence of photons.}
	\label{Fig:Model}
\end{figure}

To properly address the interference problem, we need sharper eyes to look closer and identify the light-matter interactions that rule the phenomenon. According to R. Glauber~\cite{glauber}, the interaction of the modes with the detector (matter) can be described by the Jaynes-Cummings model, whose atom-field interaction Hamiltonian is given by
$H = E^{(+)}\sigma_{+} + E^{(-)}\sigma_{-}$,
where $\sigma_+$ $\left( \sigma_- \right)$ is the Pauli raising (lowering) operator and $E^{(+)}=\sum_{m=1}^M a_me^{i\Phi_m}$ ($E^{(-)} = (E^{(+)})^{\dagger}$) is the positive (negative) frequency electric field operator, with $a_m$ ($a^\dagger_m$) being the lowering (raising) operator of the $m$-th mode in the Fock basis, with the index $m$ running over all the possible modes. The phase $\Phi_m$ can assume, in principle, any value between $[0, 2\pi)$. However, it assumes specific values either time-dependent ($\Phi_m=\varphi_m(t)=m[\Delta\omega t+\varphi]$) or time-independent ($\Phi_m=m[kd\sin(\theta)]$), in pulsed lasers and diffraction gratings, respectively. To simplify the analysis, we apply the unitary transformation  
$
U = \sum^{M}_{m=1} e^{i\Phi_m a^\dagger_m a_m}
$
which reduces the electric field operator to  
$
\Tilde{E}^{(+)} = \sum^{M}_{m=1} a_m,
$
effectively eliminating the phase dependencies. However, to maintain a precise description of the problem, the state of the entire light-matter system must also be transformed by $U$ (details in SM). With this in mind, we can construct a set of collective operators that facilitates describing the interference process. This construction is achieved by defining new operators as functions of the bare (original) ones, as~\cite{Celso_DBS}
\begin{equation}\label{collective operators}
    c_j = \sum^{M}_{m=1}O_{j,m}a_m,
\end{equation}
\noindent where $1 \leq j \leq M$ and $O$ is the change-of-basis matrix, whose $O_{1,m}=1/\sqrt{M}$, $O_{2,m}=(-1)^{m-1}/\sqrt{M}$, and the other rows are obtained by the combinations of $\pm 1/\sqrt{M}$ that satisfy the linear independence of the rows. Additionally, this particular construction of the new operators preserves the total number of photons. As better discussed in the SM, it is important to highlight that the change-of-basis matrix, as well as the subsequent results, depend on the operators forming the electric field. Thus, relative phases between these operators may also appear in the matrix for general cases, which motivates us to eliminate them at first. Provided by these collective operators, we can construct a new orthogonal basis. Since the ground state is common for both sets of operators, we can produce the other basis states as
\begin{equation}
    \ket{\Psi^N_{n_1, n_2, ..., n_M}} = \prod^{M}_{m=1}\frac{\left( c^\dagger_m \right)^{n_m}}{\sqrt{n_m !}}\ket{0,0, ..., 0}_{1', ...,M'},
\end{equation}
\noindent with the subindex $m'$ representing the $m$-th collective mode, and with  $\sum^{M}_{m=1}n_m = N$, where $N$ is the total number of photons.

As one can notice from the construction of the change-of-basis matrix, the elements of $O_{1,m}$ add over the bare operators with equal factors, leading to the symmetric collective operator $c_1$. In this way, it is the only operator that couples to matter, leading the electric field to become $\Tilde{E}^{(+)} = \sqrt{M}c_1$, which now presents an enhancement in the detector-modes coupling, which explains the higher intensity observed in the bright regions (pulses). Following this, the state where all photons can be exchanged is defined as the bright state and reads $\ket{\Psi^N_{N,0,...,0}}$. Hence, for $N$ photons, this state is unique as there is only one possible photon distribution that allows for maximum energy exchange with matter. Conversely, many combinations will result in states that do not interact with atoms, which are known as dark states. Such states are associated with antisymmetric modes, each one related to one of the remaining collective operators, in such a way that the coupling between each of them and matter is null.

The dark states are characterized by the absence of photons in the symmetric mode $(n_1=0)$, which leads to their inability to excite matter. In addition, since there are $M-1$ antisymmetric modes, several photon distributions result in dark states. The simpler case to understand how the number of dark states varies depending on the number of modes and photons is considering $M=3$ modes. In this case, for $N=1$, the dark states are $\ket{\Psi^1_{0,1,0}}$ and $\ket{\Psi^1_{0,0,1}}$ since there are just two antisymmetric modes. Now, if $N=2$, the set of dark states becomes $\left\{ \ket{\Psi^2_{0,2,0}}, \ket{\Psi^2_{0,0,2}}, \ket{\Psi^1_{0,1,1}} \right\}$, where we see one more dark state when compared to $N=1$. Considering the general case, the higher the number of modes $M$ and the total number of photons $N$, the higher the quantity of dark states forming the collective basis. The number of DS ($N_{\mathrm{DS}}$) can be obtained from
\begin{equation}\label{amount of DS general case}
    N_\mathrm{DS}=\sum^{N}_{n_2=0} \sum^{N-n_2}_{n_3=0} ... \sum^{N-n_1 - ... -n_{M-3}}_{n_{M-2}=0} \sum^{N-n_1 - ... -n_{M-2}}_{n_{M-1}=0}  1,
\end{equation}
where the absence of a sum with index $n_M$ is to preserve the total excitation because, once the number of photons in the other modes is known, the number of photons in the last mode must be the one that satisfies $\sum^{M}_{m=1}n_m = N$. 

Since we have the mathematical framework that simplifies the description of multimode systems, we will now address the light states in our analysis to start connecting the previous results to the interference problem. For this purpose, we look more carefully at single-photon~\cite{Itoh_2002, Sangbae_2023, Luo_2024} and coherent~\cite{Janszky_93, double_slit_interference_polarized_photons, young2002lectures1, young2002lectures} states, as they address fundamental concepts in the context of interference experiments. This way, the general state for a single photon equally distributed over $M$ modes is (in the Fock, or bare, basis)
\begin{equation}
\begin{split}
    &\ket{\Psi(\delta\Phi_1, ..., \delta\Phi_{M-1})}_{\text{SP}}  = \frac{1}{\sqrt{M}}\big( \ket{1,0,..., 0}_{1, ...,M} \\ &+ e^{i\delta\Phi_1}\ket{0,1, ..., 0}_{1, ...,M}  + ... + e^{i\delta\Phi_{M-1}}\ket{0,0, ..., 1}_{1, ...,M}\big),
\end{split}
\end{equation}
\noindent where, without loss of generality and aiming for mathematical rigor~\cite{book_complex_analysis}, $\delta\Phi_K \in [0, 2\pi)$, with $K=1, ..., M-1$, is the relative phase of the other modes with respect to the first one (taken as the reference mode for simplicity). The state subindex $\mathrm{SP}$ refers to the single photon case. To calculate the probability of detecting the photon (a click on the single photon detector) in this state, we apply the transformed electric field operator $\Tilde{E}^{(+)} = \sum^{M}_{m=1}a_m$, yielding
\begin{equation}\label{E single photon}
\begin{aligned}
    \Tilde{E}^{(+)} &\ket{\Psi(\delta\Phi_1, ..., \delta\Phi_{M-1})}_{\text{SP}} =\\& \left( 1 + e^{i\delta\Phi_1} + ... + e^{i\delta\Phi_{M-1}} \right)\ket{0,0, ..., 0,0}_{1, ...,M}.
\end{aligned}
\end{equation}

This equation allows us to analyze the possible dark and bright states according to the number of modes. For instance, for a single mode the coefficient reduces to $1$ and, consequently, the photon is always detected. For two modes, the coefficient becomes $\left(1 + e^{i\delta\Phi_1}\right)$, and then the state will be undetectable only if $\delta\Phi_1 = \pi$, which reduces the coefficient to zero. Interestingly, for $M=3$, the coefficient takes the form $\left(1 + e^{i\delta\Phi_1} + e^{i\delta\Phi_2} \right)$, meaning that an infinite number of relative phase sets $\left\{ \delta\Phi_1, \delta\Phi_2\right\}$ can nullify this term. Consequently, for many modes, $\ket{\Psi(\delta\Phi_1, ..., \delta\Phi_{M-1})}_{\text{SP}}$ will be undetectable for several relative phase sets. The corresponding states with these relative phases are dark states, as they do not excite the matter and are, therefore, undetectable. In contrast, for any number of modes, when all the relative phases are $\delta\Phi_K = 0$, the coefficient achieves its maximum value, indicating the highest probability of the photon being detected. Such a state is a bright one that is unique.

Similarly, for equal amplitude ($\alpha$) coherent states in $M$ modes, the general state is $ \ket{\Psi(\delta\Phi_1,..., \delta\Phi_{M-1})}_{\alpha} = \ket{\alpha, e^{i\delta\Phi_{1}}\alpha,...,  e^{i\delta\Phi_{M-1}}\alpha}_{1,...,M}$, and, when applying $\Tilde{E}^{(+)}$ we obtain 
\begin{equation}
\begin{aligned}
    \Tilde{E}^{(+)} &\ket{\Psi(\delta\Phi_1,..., \delta\Phi_{M-1})}_{\alpha} =\\& \alpha\left( 1 + e^{i\delta\Phi_1} +... + e^{i\delta\Phi_{M-1}} \right)\ket{\Psi(\delta\Phi_1,..., \delta\Phi_{M-1})}_{\alpha}, 
\end{aligned}
\end{equation}
where the coefficient is the same as the one of Eq. \eqref{E single photon}. Therefore, the previous analyses and results hold true for both states. Thus, despite their different statistics, single-photon and coherent states have the same dynamics and first-order correlations, as is well known in the literature~\cite{Walls_and_Milburn_1ed, Orszag2016}.

As we could see, without imposing any physical conditions, the relative phases of the modes can assume any value and, consequently, there is an infinite number of dark states for the multimode case (for $M\ge3$). However, the relative phases present some constraints when addressing multislit or mode-locked pulsed lasers, as previously discussed. For instance, for a single photon passing through a diffraction grating, with $M$ slits, that can be equally described by a photon emitted by entangled atoms at the positions of each slit~\cite{Scully_Zubairy_1997}, the state is $\ket{S}_\mathrm{SP}=\frac{1}{\sqrt{M}}\sum_{m=1}^{M}\bigotimes^{M}_{l=1}e^{-i\vec{k}_l \cdot \vec{d}_l \delta_{l,m}}\ket{\delta_{l,m}}_l$. Here, $\vec{d}_m$ is the position of the $m$-th slit or the equivalent atom, and $\delta_{i,j}$ is the Kronecker delta of $i$ and $j$. To apply the previous results, we must transform the $M$-mode single-photon state using the same unitary transformation $U$ that was employed to eliminate the phases in the electric field operator. In this case, the operator prior to the transformation is given by $E^{(+)}(\vec{r})=\sum_{m=1}^M a_m e^{i\vec{k}_m \cdot \vec{r}}$,  
where $\vec{r}$ represents a given position on the screen. Consequently, the phase $\Phi_m$ is defined as  
$\Phi_m = \vec{k}_m \cdot \vec{r}$,  
so that the state transforms into
\begin{equation}\label{SP state transformed}
    \ket{\Tilde{S}}_\mathrm{SP} = U \ket{S}_\mathrm{SP} 
    =\frac{1}{\sqrt{M}}\sum_{m=1}^{M}\bigotimes^{M}_{l=1}e^{i\vec{k}_l \cdot \vec{r}_l \delta_{l,m}}\ket{\delta_{l,m}}_l,
\end{equation}
where $\vec{r}_m = \vec{r} - \vec{d}_m$. By considering the slits linearly spaced, similar to the allowed frequencies in laser cavities, and taking the first mode as the reference one, the relative phases become $\Phi_m=\left|{\vec{k}_{m} \cdot \vec{r}_{m}-\vec{k}_{1} \cdot\vec{r}_{1}}\right| = (m-1)\phi$, with $\phi = kd\text{sin}(\theta)$ and  $|\vec{k}_m|=k$. Furthermore, the probability of detecting the photon in this state is obtained by substituting these linear phases in the coefficient of Eq.~\eqref{E single photon}. Hence, the coefficient can be expressed as the geometric sum $\sum^{M}_{m=1}\left(e^{i\phi}\right)^{m-1}$, whose sum of all terms yields the condition for dark states that reads (details in SM)
\begin{equation}\label{possible phis}
    \left( e^{iM\phi} -  1\right)/(e^{i\phi} - 1) = 0 \iff e^{iM\phi} = 1.
\end{equation}

Translating the above result to experimental observations in diffraction grating problems, the phase $\phi$ is related to the position where the photons hit the screen. If $\phi = 2\pi (K/M)$, with integer $K \in [1, M-1]$, $\ket{\Tilde{S}}_\mathrm{SP}$ becomes a dark state, and each $K$ yields one respective phase corresponding to one dark fringe on the screen. For $K$ outside the previous domain, i.e., when $K = 0$ or $M$, the terms in the coefficient become proportional to integer multiples of $2\pi$, in such a way the geometric sum results in $M$. These values of $K$ yield the single-photon bright states and correspond to the first or second bright fringe on the screen. Additionally, as previously discussed, the above results also hold for coherent states.

Considering specifically the case for two modes, Ref.~\cite{Celso_DBS} derives the single-photon bright and dark states as, respectively, $\ket{\Psi^1_{1,0}} = (\ket{1,0}_{1,2} + \ket{0,1}_{1,2})/\sqrt{2}$ and $\ket{\Psi^1_{0,1}} = (\ket{1,0}_{1,2} - \ket{0,1}_{1,2})/\sqrt{2}$, which are produced by the respective phases equal to $z2\pi$ $\left((2z+1)\pi\right)$, with $z$ integer. Still, it was shown that the same phases yield, respectively, the coherent bright ($\ket{\alpha, \alpha}$) and dark ($\ket{\alpha, -\alpha}$) states. This agrees with the generalized phases found here if we set $M=2$. Furthermore, in the SM, besides the detailed discussion and demonstration of the phase values that result in dark and bright states considering the linear dependency, we formally derive for $M=4$ how the single-photon state is written on the collective basis. In this case, we observe that for $\phi=z2\pi$, the state resumes to the bright state of four modes, while the phases $\phi/(2z+1)=2\pi(1/4)$, $2\pi(2/4)$, and $2\pi(3/4)$ lead to dark states, in accordance with the previous analysis.

Our findings reveal the symmetric behavior throughout the screen observed in the experiments because the results are equivalent for phases out of the range $[0, 2\pi)$. Hence, for a given reference point, $\phi=0$ and $\phi = 2\pi$ represent the first and second bright states, respectively, in which the detectable intensities are maximum, and between them there are $M-1$ dark states, whose intensities are undetectable. As the number of slits increases, so does the number of phases leading to dark states, expanding the undetectable regions, despite photons being present across the entire screen. 

The analysis above, performed for single-photon or coherent states with specific phase relationships between the modes in a diffraction grating, applies equally to mode-locked pulsed lasers, as they exhibit analogous specific phase relationships among the modes, albeit in the temporal domain. Even though photons are always present, the high number of modes producing the pulses enlarges the time windows with undetectable light. This effect, which appears from the mathematical analysis obtained from the classical description for pulsed lasers, now has, with our approach, a physical meaning derived exclusively from quantum principles and in terms of the superposition of particle (photon) states.

To check our results, we can estimate the number of modes compounding pulses using our description. As explained before, considering $M$ modes, there are $M-1$ equally spaced phases in the interval $[0, 2\pi)$ that result in dark states and only one phase that results in the bright state. As the number of modes increases, the phase separation decreases, leading to a higher density of points in dark states and forming an entirely undetectable region. Based on this, we compare the ratio of bright to dark states with the ratio of observed pulse duration to the interval between pulses. Theoretically, the duration of a pulse that presents a Gaussian shape is given by $\Delta\tau = 2 \ln(2)/\pi \Delta \nu_L$~\cite{svelto2010principles}, where $\Delta\nu_L = M \Delta\omega/2\pi$ is the total bandwidth, with $\Delta\omega$ defined after Eq.~\eqref{eq1}. Therefore, after a straightforward manipulation of the expressions, we can estimate the number of modes as (considering $n_r=1$) $M=4\ln(2) L/\pi c\Delta\tau$. For instance, a pulsed laser of $570$~ps produced in a cavity with a length equal to $53$~m, as in Ref.~\cite{Zou2020}, results in the number of modes $M=274$, approximately, yielding a ratio between bright and dark states of $\approx 1/M\sim 3.6 \times 10^{-3}$. For the same laser, the interval between the pulses experimentally observed is close to $256$~ns (corresponding to a repetition rate of $\sim 3.9$~MHz), which results in a  ``pulse''/``no pulse'' temporal ratio of ($570$~ps)/($256$~ns) $\sim 2.2 \times 10^{-3}$, matching our previous result. The difference between the ratios can be attributed to approximations made when estimating the number of modes (as discussed in the SM) and the precision of the measured pulse duration. Additionally, in the SM we apply our approach to other experimental works—some using different gain media to generate femtosecond pulses~\cite{mode_lock_optic_fiber_PRL}, and others using alternative setups to achieve specific pulse properties~\cite{Yang_25} -- which further confirms the strong agreement between theory and experiment.

In conclusion, here we studied multimodal interference, extending the new approach to describe interference phenomena from two modes in terms of superposition particle states~\cite{Celso_DBS} to an arbitrary number of modes. Our results reveal that photons are equally distributed across both the bright and dark points of the interference pattern, where the photon states are respectively coupled and uncoupled to matter. The absence of light in the dark spots is not due to a lack of photons but rather to the absence of interaction with matter. Conversely, the higher intensity observed in the bright spots does not stem from an increased number of photons but from an enhanced coupling with the detectors, which scales with the square root of the number of modes ($\sqrt{M}$), even for single-photon fields. Unlike the two-mode case, where bright and dark states always appear in pairs, the multimode case exhibits a vast number of dark states, increasing with the number of modes and photons, while the bright state remains unique. In fact, in the general case, an infinite number of dark states emerge for three or more modes. 
Exploiting the equivalence between diffraction grating and mode-locked pulsed laser, we consider the (similar) physical constraints of the problems, i.e., constant phase relationship between the modes, which reduces the number of dark states to $M-1$. Additionally, we also checked our findings by considering experimental papers regarding pulsed lasers, in which the ratio between the pulse duration (visible region) and the time separation between two consecutive pulses (dark area) aligns to the ratio between the number of bright and dark states. Our novel results may be extended to describe general pulse shapes and can pave the way for new understandings regarding the fundamentals of light-matter interaction~\cite{Dong2012, Lin2013}, information processing and erasure~\cite{luiz_universal_pc, ciro_reset_2025}, and the description of other effects originating from interference, as the ones observed in Bragg crystals~\cite{book_nonlinear_optics}.

\begin{acknowledgments}

C.M.D. and C. J. V.-B. acknowledge the support of the São Paulo Research Foundation (FAPESP, Grant No. 2022/10218-2 and No. 2022/00209-6). C. J. V.-B., B.S.S, and F.R.H.  also acknowledge the Brazilian National Council for Scientific and Technological Development (CNPq, Grant 311612/2021-0, Grant 422316/2023-7, Grant 140472/2024-0, and Grant 301976/2024-4). 

\end{acknowledgments}

\sloppy
\bibliography{bibliography.bib}

\appendix


\onecolumngrid
\newpage

\begin{center}
	{\large{ {\bf Supplemental Material for: \\ Pulsed laser as a continuous particle stream}}}

\vskip0.5\baselineskip{Ciro Micheletti~Diniz~\orcidlink{0000-0002-7602-0468}$^{1}$,  Franciele Renata Henrique~\orcidlink{0009-0000-1525-6314}$^{2}$, Bruno Santos de Souza~\orcidlink{0009-0003-9346-2844}$^{1}$, Lino Misoguti~\orcidlink{0000-0001-6624-8453}$^{2}$, Paulo Henrique Dias Ferreira~\orcidlink{0000-0003-1694-2126}$^{1}$, and Celso J. Villas~Bôas~\orcidlink{0000-0001-5622-786X}$^{1}$}






\vskip0.5\baselineskip{

   {\em $^{1}${{Departamento de Física, Universidade Federal de São Carlos, PO box 676, 13565-905 São Carlos, SP, Brazil}}}
    \\
	{\em$^{2}${{São Carlos Institute of Physics, University of São Paulo, PO Box 369, 13560-970, São Carlos, SP,
Brazil}}}
}


\end{center}

\appendix

\setcounter{equation}{0}
\setcounter{figure}{0}
\setcounter{table}{0}

\renewcommand{\theequation}{S\arabic{equation}}
\renewcommand{\thefigure}{S\arabic{figure}}
\renewcommand{\bibnumfmt}[1]{[S#1]}
\renewcommand{\citenumfont}[1]{S#1}

\appendix

\maketitle
\onecolumngrid

\section{Pulsed laser equations}
The standard (wave) explanation of the functioning of mode-locked pulsed lasers is related to the interference of the longitudinal modes of the electric field in a laser cavity.
The $m$-th mode has frequency $\omega_m = \omega_0 + [m-(M+1)/2]\Delta\omega$, where we have assumed the number of modes, M, odd for simplicity, and with $\omega_0$ being a central frequency, and $\Delta\omega =\frac{\pi c}{n_r L} \ll \text{min}(\omega_m)$ being the free spectral range for a plane mirror cavity, where $n_r$ is the refractive index of the cavity medium, $L$ is the length of the cavity, and $c$ is the speed of light in vacuum~\cite{book_laser_fundamentals_applicatons}. 
The number of modes allowed in the cavity depends on limiting factors related to geometry and, mainly, the emission bandwidth of the gain medium. In general, the classical amplitude of the output electric field (in a given direction) can be written as a sum over the allowed modes as~\cite{book_laser_fundamentals_applicatons}
\begin{equation}\label{eq1}
E(t) = \sum^{M}_{m=1}{E_m(t)e^{i\left[\omega_m t+\varphi_m(t)\right]}},
\end{equation}
\noindent where $E_m(t)$ and $\varphi_m(t)$ represent the amplitude and the temporal phase of the $m$-th mode, respectively. In principle, there is no binding between each mode's amplitude and temporal phase, and they can vary randomly according to external disturbances. However, according to O. Svelto~\cite{svelto2010principles}, we can consider $M$ ($\gg1$) modes, whose time-dependent phases are locked following the linear relation $\varphi_m(t)-\varphi_{m-1}(t) = \varphi$ (constant), and a central frequency $\omega_0\gg \Delta \omega$, which allows us to assume all $E_m(t)$ constant and equal to $E_0$. In this case,  
we are allowed to rewrite the electric field as $E(t)=A(t)e^{i[\omega_0 - (M+1)\Delta\omega/2]t}$, with the amplitude $A(t)$ given by
\begin{equation}\label{eq3}
A(t) = \sum_{m=1}^{M}{E_0e^{im(\Delta\omega t+\varphi)}}.
\end{equation}
\noindent Changing Eq.~\eqref{eq3} to another time reference in which $\Delta\omega t'=\Delta\omega t+\varphi$, the summation can be expressed as a geometric sum, resulting in
\begin{equation}\label{eq4}
A(t') \propto E_0\frac{\text{sin}\left[M\Delta\omega t'/2\right]}{\text{sin}\left[\Delta\omega t'/2\right]}.
\end{equation}
\noindent Since the laser beam intensity is proportional to $A^2(t')$, from Eq.~\eqref{eq4}, we can conclude that the interference of different modes of a laser cavity when their phases are locked results in a train of pulses, separated by the cavity's flying time $\tau=2\pi/\Delta\omega$~\cite{book_laser_fundamentals_applicatons, svelto2010principles}. Additionally, the higher the number of modes $M$ that interfere, the higher the peak intensity and the shorter the pulse duration time.

\section{Equivalence between mode-locked pulsed lasers and diffraction gratings}

In the context of interference, as shown in Fig.~\textbf{1a} of the main text, the electrical field vector operator in a given position $\vec{r}$ after passing through a diffraction grating with $M$ slits is given by, apart from an approximate constant factor, $\vec{E}^{(+)}_T = \sum^{M}_{m=1} a_m e^{i\vec{k}_m \cdot \vec{r}}\vec{e_m}$, where the annihilation operator $a_m$ represents the field mode passing through the $m$-th slit with direction given by the unit vector $\vec{e}_m$ and wavevector $\vec{k}_m$~\cite{Walls_and_Milburn_1ed}. Additionally, the interference pattern would be the same if we assume equivalently temporal phases, as observed in Ref.~\cite{Double_slit_time_interference}, where it was studied the interference pattern of a double-slit experiment considering time diffraction, thus confirming the similarity between temporal and spatial phases in interference problems. Now, if we consider a diffraction problem where the slits are linearly separated as presented in Fig.~\textbf{1} of the main text, for a light that hits the screen after passing through two consecutive slits, the difference of the optical path is given by the relation $\left|{\vec{r}_{j}-\vec{r}_{j-1}}\right| = \left|\vec{d}_{j}-\vec{d}_{j-1}\right| = d\text{sin}(\theta)$, with $d$ being the slit separation, $\theta$ is the angle between the horizontal
axis and the position vector $\vec{r}$ on a given reference point on
the screen, $\vec{d}_j$ being the position of the $j$-th slit, and $\vec{r}_j = \vec{r} - \vec{d}_j$. The angle $\theta$ can be considered constant due to the small separation between the slits when compared to the distance between the grating and the screen~\cite{double_slit_interference_geometry}. Still, by assuming a source that emits at the same frequency, we have $|\vec{k}_m|=k$, in such a way that the phases of the electrical operator can be rewritten as $e^{imkd\text{sin}(\theta)}$. This result has the same phase dependency of Eq.~\eqref{eq3}, which becomes clear since the terms $kd\text{sin}(\theta)$ and $\Delta\omega t + \varphi$ assume general values. Therefore, despite the differing physical origins -- arising from the temporal synchronization of modes in the laser versus spatial separation of slits in the diffraction grating -- the underlying physics remains analogous in both phenomena.

\section{The transformed electrical field operator}

The interaction between light and matter, considering the rotating-wave approximation (RWA), is ruled by the Hamiltonian
\begin{equation}
    H = E^{(+)}(\vec{\Phi})\sigma_{+} + E^{(-)}(\vec{\Phi})\sigma_{-},
\end{equation}
where $\sigma_+$ $\left( \sigma_- \right)$ is the Pauli raising (lowering) operator and $E^{(+)}(\vec{\Phi}) = \left( E^{(-)}(\vec{\Phi}) \right)^\dagger = \sum^{M}_{m=1} a_m e^{i\Phi_m}$, unless a constant factor, with $\Phi_m$ being general phases. Consequently, the bright and dark states of the electrical field operator $E^{(+)}(\vec{\Phi})$ also depend on the phases $\Phi_m$. In order to make the following analyses easier, we can apply a unitary transformation to eliminate these phases. 

Formally, consider $U( \vec{\Phi}) = \sum^{M}_{m=1} e^{i\Phi_m a^\dagger_m a_m}$, in such a way the electrical field operator transforms as
\begin{equation}
   \Tilde{E}^{(+)} = U( \vec{\Phi} )E^{(+)}(\vec{\Phi})U^\dagger( \vec{\Phi} ) = \sum^{M}_{m=1} a_m,
\end{equation}
\noindent and similarly for $E^{(-)}(\vec{\Phi})$. In this way, the new bright and dark states of this operator do not depend on the phases anymore. However, the Schrodinger equation that describes the evolution of the whole state of the light-matter system is transformed by $U(\vec{\Phi})$ as well. 
\begin{equation}
\begin{aligned}
    i\dot{\ket{\psi}} & = H\ket{\psi} \\ =  U^\dagger U \left(E^{(+)}(\vec{\Phi})\sigma_{+} + E^{(-)}(\vec{\Phi})\sigma_{-}\right) U^\dagger U \ket{\psi}  &= U^\dagger  \left(UE^{(+)}(\vec{\Phi})U^\dagger\sigma_{+} + UE^{(-)}(\vec{\Phi})U^\dagger\sigma_{-}\right)  U \ket{\psi} = U^\dagger \tilde{H} U \ket{\psi},
\end{aligned}
\end{equation}
\noindent where we have omitted the phase dependency in the unitary transformation for the sake of simplicity and $\tilde{H} = U ( \vec{\Phi} ) H U^\dagger (\vec{\Phi} )$. Multiplying by $U(\vec{\Phi})$ on both sides of the previous equation, we end with
\begin{equation}
    Ui\dot{\ket{\psi}} = i\dot{\left(U\ket{\psi}\right)} = \tilde{H} U \ket{\psi},
\end{equation}
where we have used that $\dot{U}\left( \vec{\Phi} \right) = 0$ once the transformation is time independent. After the calculations, the Schr\"{o}dinger equation becomes ($\hbar=1$) $i\dot{\ket{\tilde{\psi}}} = \tilde{H} \ket{\tilde{\psi}}$, where we redefine the new state and the Hamiltonian, respectively, as
\begin{subequations}
    \begin{equation}
        \ket{\tilde{\psi}} = U( \vec{\Phi} ) \ket{\psi};
    \end{equation}
    \begin{equation}
        \tilde{H} = U ( \vec{\Phi} ) H U^\dagger (\vec{\Phi} ). 
    \end{equation}
\end{subequations}

\section{Number of collective bright and dark states for a phase-fixed change-of-basis matrix}

One way to derive the bright and dark states of a collective basis is by previously knowing them on some basis. Another way to do so comes from the above discussion, exploiting the relation between the operators. In this sense, we can build collective operators as a function of the lowering operators $a_m$ of the Fock basis, also known as bare basis. Thus, we can write the collective operators as 
\begin{equation}
    c_j = \sum^{M}_{m=1} O_{j,m}a_m,
\end{equation}
\noindent where $1 \leq j \leq M$ and $O$ is the change-of-basis matrix, whose $O_{1,m}=1/\sqrt{M}$, $O_{2,m}=(-1)^{m-1}/\sqrt{M}$, and the other rows are obtained by the combinations of $\pm 1/\sqrt{M}$ that satisfy the linear independence of the rows. As an example, considering four modes we have~\cite{Celso_DBS}
\begin{equation}\label{matrix example 4 modes}
O = \begin{bmatrix} 
1/\sqrt{4} & 1/\sqrt{4} & 1/\sqrt{4} & 1/\sqrt{4} \\
1/\sqrt{4} & -1/\sqrt{4} & 1/\sqrt{4} & -1/\sqrt{4} \\
1/\sqrt{4} & 1/\sqrt{4} & -1/\sqrt{4} & -1/\sqrt{4} \\
1/\sqrt{4} & -1/\sqrt{4} & -1/\sqrt{4} & 1/\sqrt{4} 
\end{bmatrix}.
\end{equation}

This particular way of constructing the change-of-basis matrix (considering the general case with $M$ modes and the example matrix) assures the orthogonality of the operators, which, in turn, preserves canonical commutation relations and the total number of photons. The collective operator $c_1$ refers to the collective symmetric mode and the others to the collective antisymmetric modes. With this convenient set of collective operators, the electrical field operator becomes $E^{+} = \sqrt{M}c_1$. Hence, just the operator $c_1$ couples to the matter.

One can easily check that the ground state of the field operator is the same in both bases~\cite{Celso_DBS} in such a way that the states of the collective basis can be built as
\begin{equation}\label{ket colec basis}
    \ket{\Psi^N_{n_1, n_2, ..., n_{M-1}, n_M}} = \ket{n_1, n_2, ..., n_{M-1}, N - n_1 - ... - n_{M-1}}_{1', 2', ..., M'} = \prod^{M}_{m=1}\frac{\left( c^\dagger_m \right)^{n_m}}{\sqrt{n_m !}}\ket{0,0, ..., 0}_{1', 2', ..., M'},  
\end{equation}
\noindent where the subindex $m'$ represents the $m$-th collective mode, and we have the condition $\sum^{M}_{m=1}n_m = N$, and, consequently, $n_M = N - \sum^{M-1}_{m=1}n_m$.

For an arbitrary number of excitations $N$, the bright state reads $\ket{\Psi^N_{N, 0, ..., 0, 0}}$. The bright state is the one in which all the photons are in the symmetric mode and then can be exchanged. Thus, this state is unique and comes from the symmetric operator, which is capable of interacting with matter. On the other hand, the amount of dark states depends on the number of modes $M$ and the number of excitations $N$. Considering $M$ modes and $N=1$, the set of dark states is $\left\{ \ket{\Psi^1_{0,1,0,...,0,0}}, \ket{\Psi^1_{0,0,1,...,0,0}}, ..., \ket{\Psi^1_{0,0,0,...,1,0}}, \ket{\Psi^1_{0,0,0,...,0,1}} \right\}$, which means there are $M-1$ dark states in this case. However, with $N>1$, this amount increases. The simplest case to study is considering three modes. If $N=1$, there are just two dark state $\left\{ \ket{\Psi^1_{0,1,0}}, \ket{\Psi^1_{0,0,1}} \right\}$ as expect. But now, if we consider $N=2$, the set of dark states becomes $\left\{ \ket{\Psi^2_{0,2,0}}, \ket{\Psi^2_{0,0,2}}, \ket{\Psi^1_{0,1,1}} \right\}$, which means that are three dark states. Therefore, for the general case of $M$ modes and $N$ excitations, the number of dark states can be found by solving the combinatorial problem that computes the different ways to distribute $N$ photons in $M-1$ modes. This result is also the quantity of the possible values that each $n_m$ can assume while respecting the conditions $\sum^{M}_{m=1}n_m = N$ and $n_1 = 0$, which is given by
\begin{equation}
    N_\mathrm{DS} = \sum^{N}_{n_2=0} \sum^{N-n_2}_{n_3=0} ... \sum^{N-n_1 - ... -n_{M-3}}_{n_{M-2}=0} \sum^{N-n_1 - ... -n_{M-2}}_{n_{M-1}=0}  1,
\end{equation}
\noindent where we can notice the lack of the respective sum of $n_1$ because in all dark states $n_1 = 0$, and the absence of a sum with index $n_M$ because, once the number of photons in the other modes is known, the number of photons in the last mode must be the one that satisfies $\sum^{M}_{m=1}n_m = N$.

As one can notice from Eq.~\eqref{ket colec basis}, the expressions for bright and dark states depend on the collective operators, which in turn depend on the relationship between the operators of the Fock basis. Consequently, the change-of-basis matrix is not unique, and, as an example, it can be immediately seen once there are different permutations between the rows of the matrix that grant orthogonal collective operators. Additionally, the matrix and the bright and dark states expressions depend on the $E^+(\vec{\Phi})$, so the relative phases between the bare operators may also appear in the matrix. However, as shown above, we can eliminate these phases in order to simplify the analyses without losing generality.

\section{Different ways to express coherent states in four modes}

Considering the electrical field operator $\tilde{E}^{(+)}$ for two modes, it was shown in Ref.~\cite{Celso_DBS} that $\ket{\alpha,\alpha}_{1,2}= e^{-|\alpha|^{2}}\sum_{N=0}^{\infty}\sqrt{\frac{2^{N}}{N!}}\alpha^{N} \ket{\psi_{N,0}^N}$, and $\ket{\alpha,-\alpha}_{1,2} = e^{|\alpha|^{2}}\sum_{N=0}^{\infty}\sqrt{\frac{2^{N}}{N!}}\alpha^{N} \ket{\psi_{0,N}^N}$, with the $N$ photon bright and  dark states given by, respectively,
\begin{subequations}
\begin{equation}
\begin{aligned}
    \ket{\psi^N_{N,0}} & = \sqrt{\frac{N!}{2^N}}\sum^{N}_{n=0}\frac{1}{\sqrt{n!(N-n)!}}\ket{n, N-n}_{1,2},
\end{aligned}
\end{equation}
\begin{equation}
\begin{aligned}
    \ket{\psi^N_{0.N}} & =\sqrt{\frac{N!}{2^N}}\sum^{N}_{n=0}\frac{(-1)^{N-n}}{\sqrt{n!(N-n)!}}\ket{n, N-n}_{1,2}.
\end{aligned}
\end{equation}
\end{subequations}

Now, considering four modes and the respective matrix, one of the collective coherent states (taking the linguistic and definition liberties) is, for example, $\ket{\Psi}^{(M=4)}_{\alpha} = \ket{\alpha, \alpha, -\alpha, -\alpha}_{1,2,3,4}$, which is a dark state of $\tilde{E}^{(+)}$. In this way, it is straightforward that $\tilde{E}^{(+)}\ket{\Psi}^{(M=4)}_{\alpha} = \tilde{E}^{(+)}\left( \ket{\alpha, -\alpha}_{1,3} \otimes \ket{\alpha, -\alpha}_{2,4} \right) = 0$. Hence, from the previous result, it is immediate to write 
\begin{equation}\label{psi alpha 4M in two DS 2M}
     \ket{\Psi}^{(M=4)}_{\alpha} = \ket{\alpha, \alpha, -\alpha, -\alpha}_{1,2,3,4} = \ket{\alpha, -\alpha}_{1,3} \otimes \ket{\alpha, -\alpha}_{2,4}  = e^{-|\alpha|^{2}}\sum_{N'=0}^{\infty}\sqrt{\frac{2^{N'}}{N'!}}\alpha^{N'} \ket{\psi^{N'}_0}_{1,3}  \otimes e^{-|\alpha|^{2}}\sum_{\tilde{N}=0}^{\infty}\sqrt{\frac{2^{\tilde{N}}}{\tilde{N}!}}\alpha^{\tilde{N}} \ket{\psi^{\tilde{N}}_0}_{2,4}.
\end{equation}
\noindent Additionally, from the definition of coherent state, we have
\begin{equation}\label{psi alpha 4M}
\begin{split}
    \ket{\Psi}^{(M=4)}_{\alpha} & = \ket{\alpha, \alpha, -\alpha, -\alpha}_{1,2,3,4} = e^{-2 |\alpha|^2}\sum^{\infty}_{n_1 = 0} \sum^{\infty}_{n_2 = 0} \sum^{\infty}_{n_3 = 0} \sum^{\infty}_{n_4 = 0} \frac{\alpha^{n_1} \alpha^{n_2} (-\alpha)^{n_3} (-\alpha)^{n_4}}{\sqrt{n_1!n_2!n_3!n_4!}} \ket{n_1, n_2, n_3, n_4}_{1,2,3,4} \\ & = e^{-2 |\alpha|^2}\sum^{\infty}_{N = 0} \alpha^{N} \sum^{N}_{n_1 = 0} \sum^{N-n_1}_{n_2 = 0} \sum^{N-n_1-n_2}_{n_3 = 0} \frac{(-1)^{n_3} (-1)^{n_4}}{\sqrt{n_1!n_2!n_3!\left( N-n_1-n_2 - n_3 \right)!}} \ket{n_1, n_2, n_3, N-n_1-n_2 - n_3}_{1,2,3,4}.
\end{split}
\end{equation}

On the other hand, we have that the collective dark state $\ket{\Psi^N_{0, 0, N, 0}}$ in four modes is given by 
\begin{equation}
\begin{split}
    \ket{\Psi^N_{0, 0, N, 0}}  = \frac{\left(  c^\dagger_3 \right)^N}{\sqrt{N!}} \ket{0,0,0,0}_{1',2',3',4'} &= \frac{\left(  \sum^4_{m=1} O_{3,m} a^\dagger_m \right)^N}{\sqrt{N!}}\ket{0,0,0,0}_{1,2,3,4} = \sqrt{N!} \sum^{N}_{n_1 = 0} \sum^{N-n_1}_{n_2 = 0} \sum^{N-n_1-n_2}_{n_3 = 0} \prod^4_{m=1} \frac{\left( O_{3,m}a^\dagger_m\right)^{n_m}}{n_m!}\ket{0,0,0,0}_{1,2,3,4} \\&= \sqrt{N!} \sum^{N}_{n_1 = 0} \sum^{N-n_1}_{n_2 = 0} \sum^{N-n_1-n_2}_{n_3 = 0} \prod^4_{m=1} \frac{\left( O_{3,m}\right)^{n_m}}{\sqrt{n_m!}}\ket{n_1, n_2, n_3, N-n_1-n_2 - n_3}_{1,2,3,4}.
\end{split}
\end{equation}
\noindent From the change-of-basis matrix considering four modes, we have that $O_{3,1} = 1/\sqrt{4}$, $O_{3,2 }= 1/\sqrt{4}$, $O_{3,3} = -1/\sqrt{4}$, and $O_{3,4} = -1/\sqrt{4}$, in such a way that the state becomes
\begin{equation}\label{psi colec DS 3 4M}
    \ket{\Psi^N_{0, 0, N, 0}} = \sqrt{\frac{N!}{4^N}} \sum^{N}_{n_1 = 0} \sum^{N-n_1}_{n_2 = 0} \sum^{N-n_1-n_2}_{n_3 = 0}  \frac{(-1)^{n_3} (-1)^{n_4}}{\sqrt{n_1!n_2!n_3!\left( N-n_1-n_2 - n_3 \right)!}} \ket{n_1, n_2, n_3, N-n_1-n_2 - n_3}_{1,2,3,4}.
\end{equation}
\noindent Hence, by comparing Eqs.~\eqref{psi alpha 4M} and \eqref{psi colec DS 3 4M}, we can write 
\begin{equation}\label{psi alpha 4M in one DS 4M}
    \ket{\Psi}^{(M=4)}_{\alpha} = e^{-2 |\alpha|^2}\sum^{\infty}_{N = 0} \alpha^{N} \sqrt{\frac{4^N}{N!}}  \ket{\Psi^N_{0, 0, N, 0}},
\end{equation}
\noindent which is a third alternative way to express the four-modes-coherent state $\ket{\alpha, \alpha, -\alpha, -\alpha}_{1,2,3,4}$. Throughout equivalent derivations, one can obtain similar results presented here for the collective bright state and the other dark states. In addition, looking at Eqs.~\eqref{psi alpha 4M in two DS 2M} and \eqref{psi alpha 4M in one DS 4M}, we notice their similarity, which suggests one can decompose the four-mode collective state $\ket{\Psi^N_{0, 0, N, 0}}$ as a tensor product of the two-mode collective states $\ket{\Psi^N_{0, N}}_{1,3}$ and $\ket{\Psi^N_{0, N}}_{2,4}$. Furthermore, this result indicates that any multi-mode collective states could be written as tensor products of collective states of a lesser number of modes. The decomposition of a specific collective state into collective states from smaller subspaces may depend on the state undergoing decomposition and the states of the collective bases. For example, in the case treated above, the decomposition takes just two collective states from smaller subspaces, both with the same dimension, and it also seems to be symmetric, i.e., the weights of the tensor product might be the same for both states. However, for general cases, it is possible that the decomposition needs more than two collective states and not necessarily with the same dimensions or still with different weights, probably depending on the number of collective operators and the phase relationships present in the state undergoing the decomposition and in the collective basis of the smaller subspaces.

\section{The equivalence between the results for collective single-photon and coherent states}

The general state for single photons considering $M$ modes is given by 
\begin{equation}
    \ket{\Psi(\delta\Phi_1, \delta\Phi_2, ..., \delta\Phi_{M-1})}_{\text{SP}} = \frac{1}{\sqrt{M}}\left( \ket{1,0,0,..., 0}_{1,2, ...,M} + e^{i\delta\Phi_1}\ket{0,1,0, ..., 0}_{1,2, ...,M} + ... + e^{i\delta\Phi_{M-1}}\ket{0,0,0, ..., 1}_{1,2, ...,M}\right),
\end{equation}
\noindent or in a more compact way
\begin{equation}
    \ket{\Psi(\delta\Phi_1, \delta\Phi_2, ..., \delta\Phi_{M-1})}_{\text{SP}} = \frac{1}{\sqrt{M}}\left(\ket{1,0,0,..., 0}_{1,2, ...,M} + \sum_{m=2}^{M}\bigotimes^{M}_{l=1}e^{i\delta\Phi_{l-1} \delta_{l,m}}\ket{\delta_{l,m}}_l\right),
\end{equation}
\noindent where $\delta\Phi_K \in [0, 2\pi)$, with $K=1, ..., M-1$. In the same way, the general coherent state considering $M$ modes reads
\begin{equation}
    \ket{\Psi(\delta\Phi_1, \delta\Phi_2, ..., \delta\Phi_{M-1})}_{\alpha} = \ket{\alpha, e^{i\delta\Phi_{1}}\alpha, e^{i\delta\Phi_{2}}\alpha,...,  e^{i\delta\Phi_{M-1}}\alpha}_{1,2, ...,M}.
\end{equation}

Considering general phases $\delta\Phi_K$, we can apply the transformed electrical field operator $\tilde{E}^{(+)} = \sum^{M-1}_{K=1}a_K$ in both states, yielding
\begin{subequations}\label{E nos Phi}
    \begin{equation}\label{E nos Phi a}
         \tilde{E}^{(+)}\ket{\Psi(\delta\Phi_1, \delta\Phi_2, ..., \delta\Phi_{M-1})}_{\text{SP}} = \left( 1 + e^{i\delta\Phi_{1}} + e^{i\delta\Phi_{2}} + ... + e^{i\delta\Phi_{M-1}} \right) \ket{0,0,0, ..., 0}_{1,2, ...,M},
    \end{equation}
    \begin{equation}\label{E nos Phi b}
        \tilde{E}^{(+)}\ket{\Psi(\delta\Phi_1, \delta\Phi_2, ..., \delta\Phi_{M-1})}_{\alpha} = \alpha\left( 1 + e^{i\delta\Phi_{1}} + e^{i\delta\Phi_{2}} + ... + e^{i\delta\Phi{M-1}} \right) \ket{\alpha, e^{i\delta\Phi_{1}}\alpha, e^{i\delta\Phi_{2}}, ..., e^{i\delta\Phi_{M-1}}\alpha}_{1,2, ...,M}.
    \end{equation}
\end{subequations}

\noindent At first, it is worth noticing the equivalence between the results for coherent and single-photon cases revealed by the coefficients in Eq.~\eqref{E nos Phi}. Despite the different physical characteristics of these states, if we assume a given set of phases, the results observed, i.e., when the state is measured, must be the same. Therefore, once we do not cite or use a specific kind of state and its properties to derive the results, the previous and subsequent analyses must be similar for both of them.

Another important highlight observed for both cases is the number of bright and dark states that can be found for an arbitrary set of phases. Considering $M=2$ modes, the coefficient when applying the electrical field operator resumes to $\left( 1 + e^{i\delta\Phi_{1}} \right)$, in such a way that the phase that produces the unique bright state $\left( \ket{ \Psi(\delta\Phi_1)_B} \propto \ket{1,0}_{1,2} + \ket{0,1}_{1,2} \right) $ is $\delta\Phi_1 = 0$. Conversely, the phase that produces the unique dark state $\left( \ket{ \Psi(\delta\Phi_1)_D} \propto \ket{1,0}_{1,2} - \ket{0,1}_{1,2} \right)$ is $\delta\Phi_1 = \pi$. Going further and considering $M > 2$, the set of phases that produces the bright state is unique and is given by $\delta\Phi_K = 0~\forall~K$, resulting in the maximum value for the coefficient, which is $M$. On the other hand, infinite sets of phases produce dark states since every $\delta\Phi_K$ can assume any value in the continuous trigonometric circle. For example, considering $M=3$ modes, the coefficient when applying the electrical field operator reads $\left( 1 + e^{i\delta\Phi_{1}} + e^{i\delta\Phi_{2}} \right)$, where it is clear that $e^{i\delta\Phi_{1}} + e^{i\delta\Phi_{2}} = -1 $ for infinite sets of $\{ \delta\Phi_{1}, \delta\Phi_{2}  \}$.

Looking back to the problem addressed in the Letter, which is the multimodal interference, the phases in the many slits interference and the pulsed laser are bounded to each other due to the physics of the problems, either obeying the linear relation $\Phi_m = (m - 1)\phi$, where $\phi = kd \sin{\theta}$ in the former problem and  $\phi = \Delta \omega t + \varphi$ in the latter. Thus, once the phase $\phi$ is locked, all the phases $\Phi_m$ become determined, and the whole coefficient in Eqs.~\eqref{E nos Phi a} and \eqref{E nos Phi b} now depends only on the phase $\phi$. In fact, we can conveniently rewrite it as 
\begin{equation}\label{coef phase locked}
    1 + e^{i\phi} + e^{i2\phi} + ... + e^{i(M-2)\phi} + e^{i(M-1)\phi} = \sum^{M}_{m=1} \left( e^{i\phi} \right)^{m-1},
\end{equation}
\noindent which corresponds to the geometric sum, in which the first term is $1$ and the ratio is $q=e^{i\phi}$. Thus, we can compute the value of the coefficient by calculating the sum of the $M$ terms. This sum reduces simply to $\left( e^{iM\phi} -  1\right)/(e^{i\phi} - 1)$. Hence, the dark states are found by making the result equal to zero, which means the respective phases lead to states that have zero probability of being detected. This yields the condition
\begin{equation}\label{possible phis}
    \left( e^{iM\phi} -  1\right)/(e^{i\phi} - 1) = 0 \iff e^{iM\phi} = 1.
\end{equation}
\noindent Since we can write $1=e^{i2z\pi}$ with $z$ integer, the phases that satisfy the Eq.~\eqref{possible phis} are $\phi = 2\pi(z/M)$. Therefore, we can determine all the phases that result in dark states. Additionally, $z$ must be smaller than $M$ to ensure the validity of the exponential as a single-valued function~\cite{book_complex_analysis}. Still, without loss of generality due to the symmetry of the problem, we have already anticipated this issue and circumvented it by properly defining the phases $\delta\Phi_K$ in the set $[0, 2\pi)$. When $z$ is an integer multiple of $M$, the phase $\phi$ becomes solely a multiple of $2\pi$, which yields bright states. As a last remark, the coefficient of Eq.~\eqref{E nos Phi} is general and holds for any set of phases $\Phi_k$. Hence, for any other phase relationship (not necessarily the linear relation), one may be able to find, at least numerically, the phases that result in bright and dark states.


\section{Collective single-photon state in four modes}

Let's first examine the diffraction grating problem with four slits without imposing any physical restrictions between them. Hence, for this case, the electrical field operator at a given position $\vec{r}$ is $E^{(+)}(\vec{r}) = \sum_{m=1}^4a_me^{i\vec{k}_m\cdot\vec{r}}$, and following the approach of Ref. \cite{Scully_Zubairy_1997}, the single photon state emitted by entangled atoms at the slit positions $\vec{d}_m$ reads
\begin{align}
\begin{split}
    \ket{S}_4 = \frac{1}{2}\sum_{m=1}^{4}e^{-i\vec{k}_m \cdot \vec{d}_m}\ket{m},
\end{split}
\end{align}
with the adopted convention $\ket{1}=\ket{1,0,0,0}_{1,2,3,4}$ to refer to the single photon being in the first mode (considering the Fock basis), and similarly to the other states, \textit{e.g.}, $\ket{4}=\ket{0,0,0,1}_{1,2,3,4}$. Applying the unitary transformation $U = \sum^{M}_{m=1} e^{i(\vec{k}_m \cdot \vec{r}) a^\dagger_m a_m}$ in order to eliminate the phases present in the electric field, and accordingly transform the previous state in order to exploit the previous results, we obtain
\begin{align}\label{State 4 slits}
\begin{split}
    \ket{\tilde{S}}_4 = \frac{1}{2}\sum_{m=1}^{4}e^{i\vec{k}_m \cdot \vec{r}_m}\ket{m},
\end{split}
\end{align}
where $\vec{r}_m = \vec{r} - \vec{d}_m$. The previous state in Eq. \eqref{State 4 slits} is expressed using the Fock basis; however, we can consider the problem via the collective-mode basis in order to deal with it similarly as in Ref. \cite{Celso_DBS}. In this sense, the change-of-basis matrix arises from the relation between the operators $a_m$ of the transformed electrical field and the collective operators of the symmetric and antisymmetric modes. Following our results, for four modes, the collective operators reduce to $c_j = \sum_m O_{j,m} a_m$, where $O$ is the change-of-basis matrix of Eq.~\eqref{matrix example 4 modes}. The collective operator $c_1$ refers to the symmetric mode and the other ones to the antisymmetric modes. Hence, it is possible to rewrite the Fock states as functions of the collective (bright and dark) states in such a way that the state of Eq.~\eqref{State 4 slits} becomes
\begin{align}\label{State 4 slits colec basis}
\begin{split}
    \ket{\tilde{S}}_4 = \frac{1}{4}\Bigg(\left(e^{i\vec{k}_1\cdot\vec{r}_1} + e^{i\vec{k}_2\cdot\vec{r}_2} + e^{i\vec{k}_3\cdot\vec{r}_3} + e^{i\vec{k}_4\cdot\vec{r}_4}\right)\ket{1'} \\ + \left(e^{i\vec{k}_1\cdot\vec{r}_1} - e^{i\vec{k}_2.\vec{r}_2} + e^{i\vec{k}_3\cdot\vec{r}_3} - e^{i\vec{k}_4\cdot\vec{r}_4}\right)\ket{2'} \\
    + \left(e^{i\vec{k}_1\cdot\vec{r}_1} + e^{i\vec{k}_2\cdot\vec{r}_2} - e^{i\vec{k}_3\cdot\vec{r}_3} - e^{i\vec{k}_4\cdot\vec{r}_4}\right)\ket{3'} \\ + \left(e^{i\vec{k}_1\cdot\vec{r}_1} - e^{i\vec{k}_2\cdot\vec{r}_2} - e^{i\vec{k}_3\cdot\vec{r}_3} + e^{i\vec{k}_4\cdot\vec{r}_4}\right)\ket{4'}\Bigg),
\end{split}
\end{align}
where we have defined $\ket{j'}=\sum_{m=1}^4 O_{j,m}\ket{m}$, with $1\leq j'\leq4$. Now, if we assume a diffraction problem obeying the linear relation $\left|{\vec{k}_{m} \cdot \vec{r}_{m}-\vec{k}_{1} \cdot\vec{r}_{1}}\right| = (m-1)kd\text{sin}(\theta)$, as presented in Fig.~\textbf{1a}, the phases of the previous state become proportional to a general phase $\phi=kd\sin(\theta)$. In this sense, employing the previous relationship and after some direct labor manipulating the exponentials, we can rewrite the previous state as (apart from a global phase)
\begin{align}\label{State 4 slits colec basis derived}
\begin{split}
    \ket{S}_4 = \text{cos}\left(\phi/2\right)\text{cos}\left(\phi\right)\ket{1'}  - i\text{sin}\left(\phi/2\right)\text{cos}\left(\phi\right)\ket{2'}  
   \\  + i\text{cos}\left(\phi/2\right)\text{sin}\left(\phi\right)\ket{3'} - \text{sin}\left(\phi/2\right)\text{sin}\left(\phi\right)\ket{4'},
\end{split}
\end{align}
\noindent If $\phi/2=2z\pi$, with $z$ integer, the resulting state is $\ket{1'}$, which is the bright state. On the other hand, if $(\phi/2)/(2z + 1)=\pi/2$, $\pi/4$, or $3\pi/4$, $\ket{S}_4$ results in one of the dark state ($\ket{2'}$, $\ket{3'}$, $\ket{4'}$) or a combination of some them.

\section{Mode-locked pulsed laser: comparison between experimental results and theoretical valuations}

We can estimate the number of modes compounding some pulses using our description. Considering $M$ modes, there are $M-1$ equally spaced phases in the interval $[0, 2\pi)$ that result in dark states and only one phase that results in the bright state. As the number of modes increases, the phase separation decreases, leading to a higher density of points in dark states and forming an entirely undetectable region. Based on this, we can compare the theoretical ratio of bright to dark states with the ratio of experimentally observed pulse duration to the interval between pulses. Following the laser theory, and in order to simplify our qualitative analyses, let's consider a Gaussian shape pulse in which the time duration is given by $\Delta\tau = 2 \ln(2)/\pi \Delta \nu_L$~\cite{svelto2010principles}, where $\Delta\nu_L = M \Delta\omega/2\pi$ is the total oscillating bandwidth, with $\Delta\omega =\frac{\pi c}{n_r L}$. Therefore, after a straightforward manipulation of the expressions, we can estimate the number of modes as (considering $n_r=1$) $M=4 \ln(2) L/\pi c\Delta\tau$. Additionally, the time separation between the pulses can be estimated as $\tau=2\pi/\Delta\omega = 2n_rL/c$~\cite{book_laser_fundamentals_applicatons, svelto2010principles}, in such a way that the theoretical ratio of bright to dark states is given by $\approx 1/M$, and the experimental ratio of the pulse duration to the time interval between them reads $\Delta\tau/\tau$. With these in hand, in principle, we can check the results we derived here for any arbitrary case that meets our considerations. Here, we present some references in which we compare the theoretical and experimental ratios to put our results to the test. 

Starting by Ref.~\cite{Zou2020}, where a pulsed laser of $570$~ps is produced in a cavity with a length equal to $53$~m, the resulting number of modes is $M\sim274$, approximately, yielding a ratio between bright and dark states of $\approx 1/M\sim 3.6 \times 10^{-3}$. For the same laser, the interval between the pulses experimentally observed is close to $256$~ns (corresponding to a repetition rate of $\sim 3.9$~MHz), which results in a  “light”/“no light” ratio of ($570$~ps)/($256$~ns) $\sim 2.2 \times 10^{-3}$, matching our previous result. As another example, two lasers are considered in first part of the results of Ref.~\cite{Yang_25}, one with a duration of $577.83$~ps and repetition rate of $3.18$~MHz (corresponding to a cavity of $62.89$~m), and the other with a duration of $343.41$~ps and repetition rate of $2.62$~MHz (corresponding to a cavity of $76.34$~m). The theoretical (experimental) ratios are, respectively, $\sim 3.1 \times 10^{-3}$ ($\sim 1.8 \times 10^{-3}$) for the first case and $\sim 1.5 \times 10^{-3}$ ($\sim 0.9 \times 10^{-3}$) for the second case, showing again great alignment between theory and experiment. Still, in Ref.~\cite{mode_lock_optic_fiber_PRL}, a femtosecond pulsed laser ($275.5$~fs) with a repetition rate of $120$~MHz (corresponding theoretically to a cavity length of $1.25$~m) is produced, yielding a ratio of bright to dark states of $\approx 7.5 \times 10^{-3}$, and a ratio of pulse-to-interval duration pulse of $\approx 3.3 \times 10^{-3}$.

As an example of caution, in Ref.~\cite{Wang_12_TiSa_laser}, a laser of $16$~ns with a repetition rate of $1$~KHz is reported, where a cavity of $250$~mm is used. In this study, additional dispersion components are used to compress the pulse, which makes the two ratios different since the shape of the pulse is modified. This modification after the pulse generation means that the number of modes compounding the pulse does not match the pulse characteristics. In fact, if we try to estimate the number of modes by considering only the previous information, we obtain a number of modes smaller than unity, which has no physical meaning. Additionally, the theoretical time separation ($\tau \sim 1.7 \times 10^{-9}$~s) is very different from the experimental observation ($ \approx 1/(1\text{KHz}) = 10^{-3}$~s), showing that additional manipulations are done to achieve the pulses reported.

In conclusion, considering the cases in which both the expressions employed and the conditions adopted in their derivations are valid, we achieve a great alignment between our theoretical results and experimental observations. The difference between the ratios can be attributed to errors in approximations while estimating the number of modes, such as the ones arising from experimental uncertainty or still from the description of the problem since the mathematical expressions have different factors depending on the characteristics of the systems and pulses, which can include the type of the cavity, where the shutter is placed, and the shape of the pulse~\cite{svelto2010principles, mode_lock_random_laser_nature_communi} (\textit{e.g.}, for pulses that have gaussian shape, the factor is $2\ln{2}/\pi \sim 0.441$. For other pulses, as the ones approximated by hyperbolic-secant-squared shapes, this factor is different). Also, the presence of the \textit{intermediate states}, which are neither bright nor dark states but play a role in the pulse shape~\cite{Celso_DBS}, can also trigger approximation errors.

\end{document}